\documentclass[twocolumn,appendixfloats]{aastex62}
\submitjournal{ApJS}
\shorttitle{2MRS in the ZoA}
\shortauthors{Macri et al.}
\newcommand{\ntot}{44,572}
\newcommand{\ngal}{1,041}
\begin{document}

\title{\sc The 2MASS Redshift Survey in the Zone of Avoidance}

\correspondingauthor{Lucas M.~Macri}
\email{lmacri@tamu.edu}

\author[0000-0002-1775-4859]{Lucas M.~Macri}
\affil{George P. \& Cynthia Woods Mitchell Institute for Fundamental Physics and Astronomy, Department of Physics \& Astronomy,\\Texas A\&M University, 4242 TAMU, College Station, TX 77843-4242, USA}

\author[0000-0002-0202-6250]{Ren\'ee~C.~Kraan-Korteweg}
\affiliation{Department of Astronomy, University of Cape Town, Private Bag X3, Rondebosch 7701, South Africa}

\author[0000-0001-6263-0970]{Trystan~Lambert}
\affiliation{South African Astronomical Observatory, PO Box 9, Observatory 7935, Cape Town, South Africa}

\author[0000-0001-9684-589X]{Mar\'{\i}a Victoria Alonso}
\affiliation{Instituto de Astronom{\'\i}a Te\'orica y Experimental (IATE-CONICET), Laprida 854, X5000BGR C\'ordoba, Argentina}
\affiliation{Observatorio Astron\'omico de C\'ordoba Universidad Nacional de C\'ordoba, C\'ordoba, Argentina}

\author{Perry~Berlind}
\affiliation{Harvard-Smithsonian Center for Astrophysics, 60 Garden St, Cambridge, MA 02138, USA}

\author[0000-0002-2830-5661]{Michael~Calkins}
\affiliation{Harvard-Smithsonian Center for Astrophysics, 60 Garden St, Cambridge, MA 02138, USA}

\author[0000-0002-3430-3822]{Pirin~Erdo{\u g}du}
\affiliation{Department of Physics and Astronomy, University College London, Gower St, London WC1E 6BT, UK}

\author[0000-0002-7061-6519]{Emilio E.~Falco}
\affiliation{Harvard-Smithsonian Center for Astrophysics, 60 Garden St, Cambridge, MA 02138, USA}

\author[0000-0002-4939-734X]{Thomas~H.~Jarrett}
\affiliation{Department of Astronomy, University of Cape Town, Private Bag X3, Rondebosch 7701, South Africa}

\author[0000-0003-3594-1823]{Jessica~D.~Mink}
\affiliation{Harvard-Smithsonian Center for Astrophysics, 60 Garden St, Cambridge, MA 02138, USA}

\begin{abstract}
The 2MASS Redshift Survey was started two decades ago with the goal of mapping the three-dimensional distribution of an all-sky flux-limited ($K_s<11.75$~mag) sample of $\sim 45,000$~galaxies. Our first data release \citep{huchra12} presented an unprecedented uniform coverage for most of the celestial sphere, with redshifts for $\sim 98$\% of our sample. However, we were missing redshifts for $\sim 18$\% of the catalog entries that were located within the ``Zone of Avoidance'' ($|b|<10^\circ$) -- an important region of the sky for studies of large-scale structure and cosmic flows.

In this second and final data release, we present redshifts for all {\ngal} 2MRS galaxies that previously lacked this information, as well as updated measurements for 27 others.
\end{abstract}


\section{Introduction} \label{sec:intro}

Decades of concerted efforts have yet to resolve the tension between local cosmic flows and cosmology: what gives rise to our peculiar motion, and is it consistent with $\Lambda$CDM? Answering these question requires knowledge of the local cosmography (based on the observed distribution of luminous matter) and the peculiar velocity field (which originates from the combined distribution of both luminous and dark matter). The persistent discrepancies in the derived Local Group motion are thought to be due to incomplete mapping of large-scale structures at low Galactic latitudes, in the so-called ``Zone of Avoidance'' \citep[ZoA;][]{kraan00,loeb08}.

Obtaining a complete redshift sample in the ZoA is crucial for this endeavor. This area of the sky contains some of the largest mass concentrations in the local Universe, such as the core of the Great Attractor \citep{kraan96}, the Pisces-Perseus Supercluster \citep{ramatsoku14,ramatsoku16}, and the recently-discovered Vela Supercluster \citep{kraan17}. This latest discovery highlights the importance of spectroscopic surveys for finding and mapping additional large mass concentrations.

Our understanding of the large-scale structures and associated dynamics in the nearby Universe is increasingly being refined thanks to the advances of larger and more systematic redshift surveys. The Two Micron All-Sky Survey (2MASS) produced an ``extended source catalog'' with $\sim1.5\times 10^6$ galaxies complete to $K_S\!=\!13.5$ \citep{jarrett00}. The 2MASS Redshift Survey \citep[2MRS:][]{huchra05} was started two decades ago with the goal of obtaining redshifts for \underline{all} of the $\sim 45,000$ brightest 2MASS galaxies ($K_s<11.75$, extinction-corrected) outside of the innermost ZoA (defined as $|b|\!<\!8^\circ$ for $-30^\circ\!<\!l\!<\!30^\circ$ and $|b|\!<\!5^\circ$ otherwise).

\begin{figure*}[t]
 \includegraphics[width=\textwidth,angle=0]{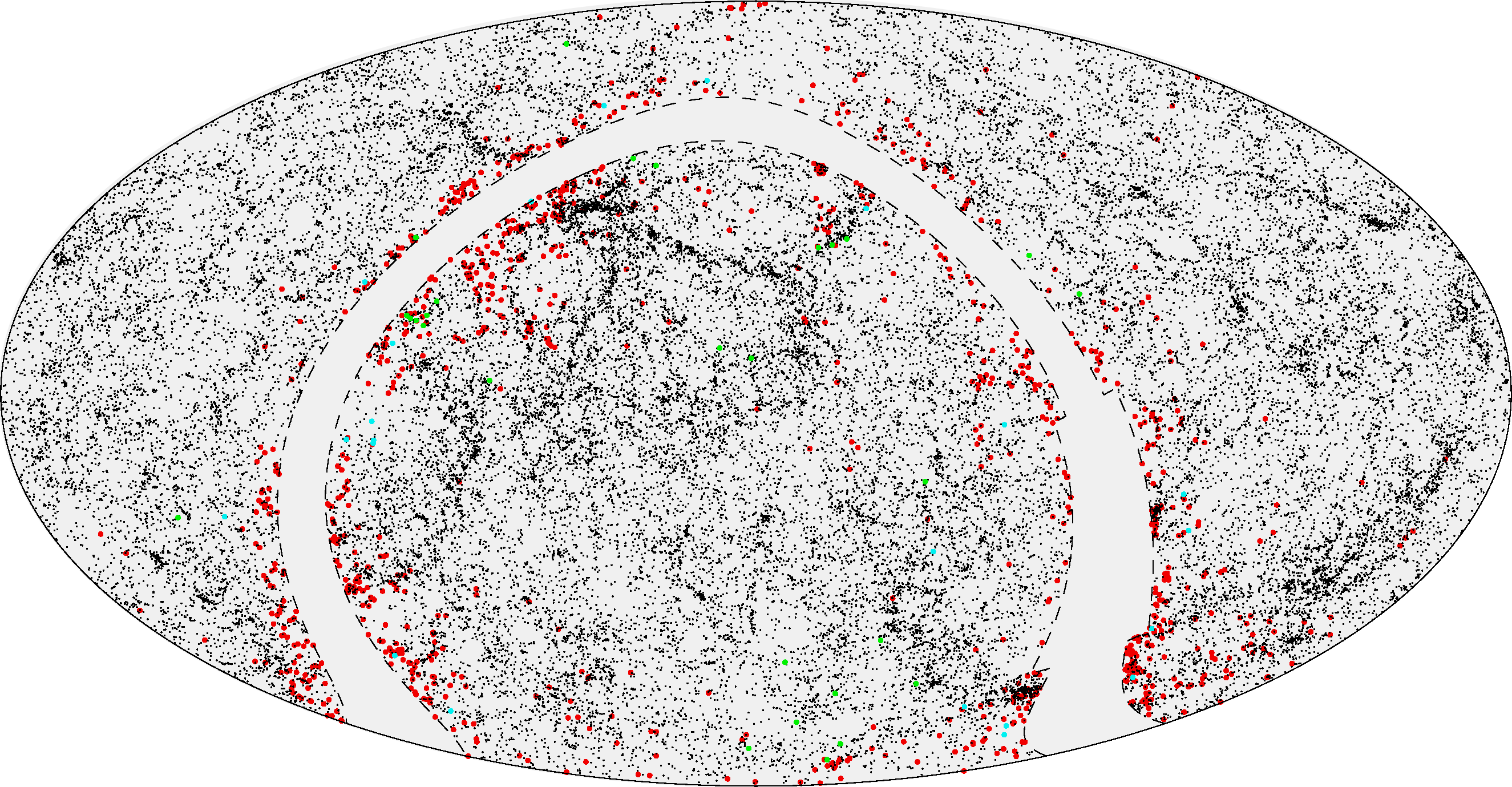}\\
 \includegraphics[width=\textwidth,angle=0]{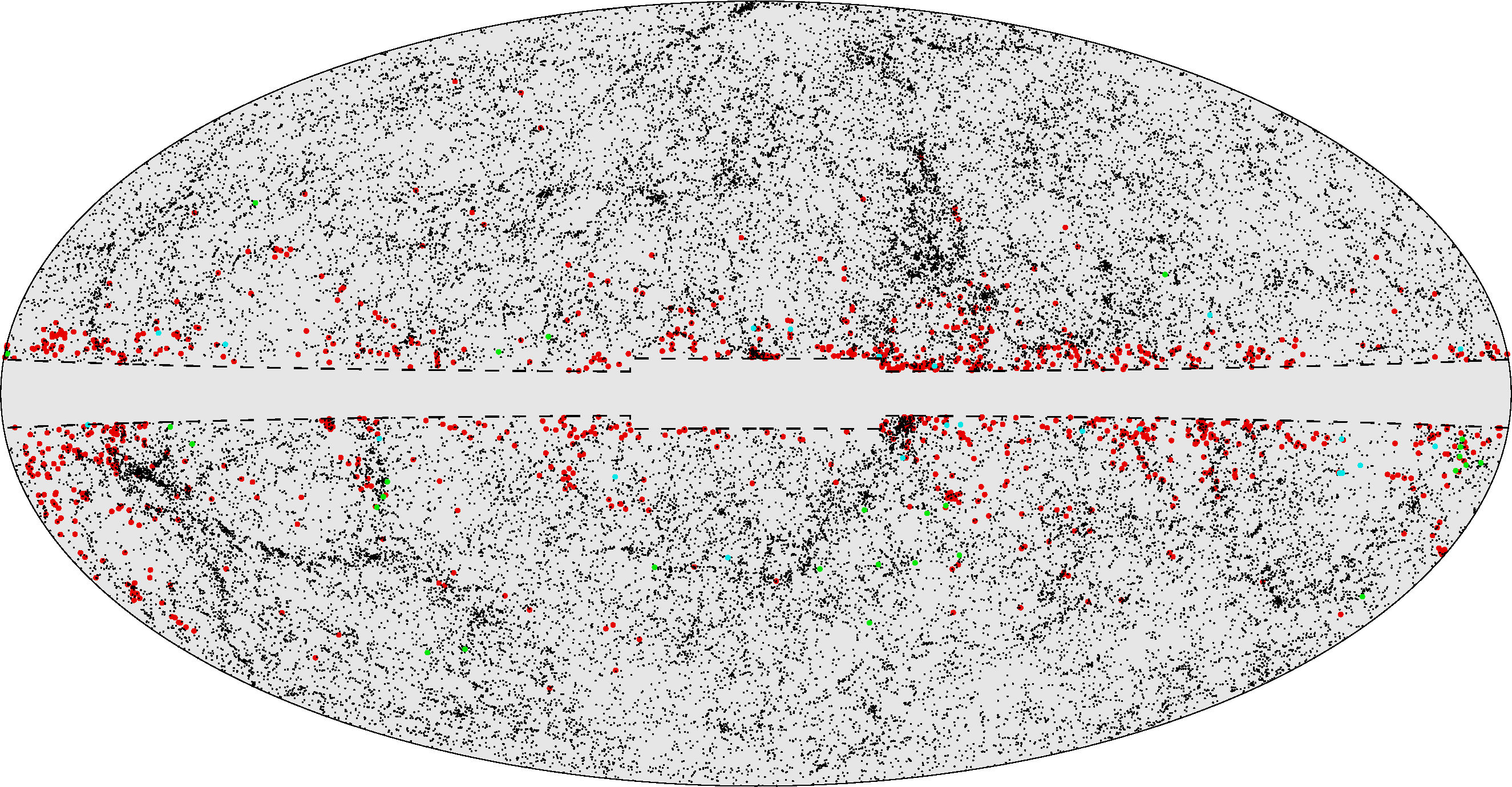}
\caption{All-sky distribution of 2MRS galaxies in equatorial (top) and Galactic (bottom) coordinates. These Aitoff projections are centered at $(\alpha,\delta)$ and $(l,b)$ of $(0,0)$ with the longitudinal coordinate increasing to the left. Black dots: galaxies with redshifts presented in H12; red dots: galaxies without redshifts in H12, presented here for the first time; green dots: galaxies with redshifts in H12 that have been updated; cyan dots: entries removed from the catalog (mistakenly classified as galaxies). Dashed lines indicate the boundaries of 2MRS ($|b|<8^\circ$ for $-30^\circ < l < 30^\circ; |b|<5^\circ$ otherwise).\label{fig:allsky}}
\end{figure*}

\vfill\pagebreak\newpage

Our initial data release \citep[][hereafter H12]{huchra12} presented measurements for 97.6\% of the sample, including new observations of 11,000 galaxies with previously unknown redshifts. 2MRS has been (and continues to be) extensively used for studies of the large-scale structure \citep{jarrett04}, group and clustering properties \citep[][T.~Lambert et al., in preparation]{crook07}, the dipole motion \citep{erdogdu06a}, and the local density and velocity fields \citep{erdogdu06b,hess16}. The 2MASS Tully-Fisher survey \citep[2MTF;][]{masters08} obtained high-quality 21-cm linewidths for 493 2MRS galaxies which, combined with measurements from the literature, have yielded a catalog of 2062 distances based on the Tully-Fisher relation \citep{tully77} and corresponding peculiar velocities \citep{hong19}. Among other applications, 2MTF has been used to study the velocity power spectrum \citep{howlett17} and to determine the bulk flow out to $D\sim 40$~Mpc \citep{hong19}. These efforts are complementary to those of the 6dFGS project \citep{jones09}, which obtained redshifts for a deeper ($K_s \lesssim 12.5$~mag) sample of 2MASS galaxies (but was limited to the Southern hemisphere) and used the fundamental plane of elliptical galaxies \citep{dressler87} to derive distances \citep{magoulas12}.

\section{Observations} \label{sec:obs}

Our previous data release (H12) consisted of 43,533 and 1,066 galaxies with and without redshifts, respectively. During the course of our observations, we identified 26 objects in the latter set that were removed from the catalog; these were stars, badly-centered duplicates of other galaxies already in the catalog, or star/galaxy blends that severely compromised the photometry of the extended source. Additionally, we removed one galaxy from the set with redshifts, which was a badly-centered duplicate of an object present in the set without redshifts. The existing redshift measurement was reassigned to the latter. After these changes, the 2MRS catalog consists of \ntot\ galaxies (please refer to Table~\ref{tab:chg} for details of all changes relative to H12). Fig.~\ref{fig:allsky} shows the all-sky distribution of 2MRS sources in equatorial and Galactic coordinates.

We noticed that two galaxies ($08533956\!+\!7256532$ and $17564915\!-\!8156263$) had been erroneously assigned redshifts in H12 so they were moved to the set without redshifts, which consisted of \ngal\ targets after all the aforementioned changes. One of us (R.K.K.) contributed previously-unpublished measurements for 39 objects and we found redshifts in NED (NASA/IPAC Extragalactic Database) for five galaxies \citep{crook07,guzzo09,schroder09}, leaving 997 objects to be observed.

Given the relatively bright apparent magnitudes ($V\lesssim 16$~mag) and the low density of our targets in the sky ($\sim 1$~per sq.~deg.), observations of these objects are most efficiently carried out using medium-aperture ($2-4$~m) telescopes with long-slit spectrographs. We used the following facilities between 2012 May and 2018 September to obtain 1,023 redshifts (for the 997 aforementioned targets and for repeat observations of 26 galaxies with redshifts in H12, mainly due to very large uncertainties in the previous measurements):

\begin{itemize}

\item The Tillinghast 1.5-m telescope at the Fred L.~Whipple Observatory (FLWO) with the ``FAst Spectrograph for the Tillinghast'' \citep[FAST;][]{fabricant98} yielded 573 redshifts (code ``F'' on Table~\ref{tb:cat}).

\item The Radcliffe 1.9-m telescope at the South African Astronomical Observatory (SAAO) with the CassSpec and ``Spectrograph Upgrade: Newly Improved Cassegrain'' (SpUpNIC) spectrographs \citep{crause19} yielded 403 redshifts (code ``Z'' on Table~\ref{tb:cat}).

\item The Jorge Sahade 2.2-m telescope at the Complejo Astron\'omico El Leoncito (CASLEO) with the REOSC spectrograph \citep{baume17} yielded 37 redshifts (code ``L'' on Table~\ref{tb:cat}).

\item The Southern Astrophysical Research (SOAR) 4.1-m telescope at the Cerro Tololo Interamerican Observatory with the Goodman spectrograph \cite{clemens04} yielded 8 redshifts (code ``G'' on Table~\ref{tb:cat}).

\item The Hiltner 2.4-m telescope at the Michigan-Dartmouth-MIT (MDM) observatory with the Ohio State Multi-Object Spectrograph \citep[OSMOS;][]{martini11} yielded 2 redshifts (code ``M'' on Table~\ref{tb:cat}).
\end{itemize}

\begin{figure*}[t]
\begin{center}
\includegraphics[width=0.9\textwidth,angle=0]{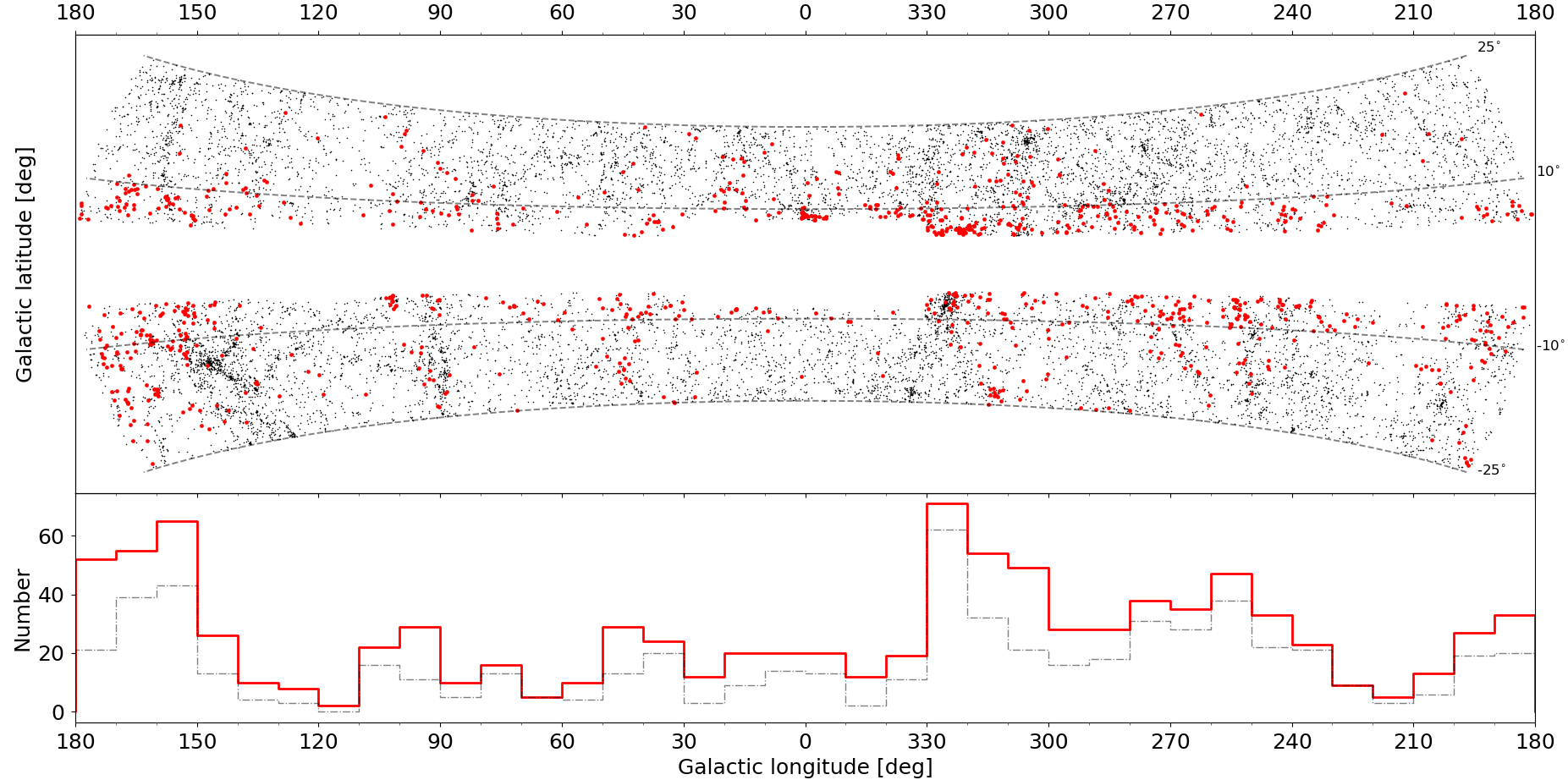}
\end{center}
\caption{Distribution of the 1,068 new redshifts that complete 2MRS. Top: zoom of Fig.~1b, focusing on $|b| \leq 25^{\circ}$; red dots (marginally enlarged for visibility) represent the new redshifts, while black dots show the galaxies from H12. Bottom: histogram of the new redshifts as function of Galactic longitude for $|b| \leq 25^{\circ}$ (red line) and the inner ZOA $|b| \leq 10^{\circ}$ (dashed grey line).\label{fig:zoa}}
\end{figure*}

\section{Analysis and Results} \label{sec:results}

All spectra were analyzed following the same procedures described in detail in \S3.2 of H12. Our measurements have been corrected to the barycentric reference frame, and all corrections were applied using the proper product of $(1+z)$ factors. We chose to express our redshifts in units of $cz$, noting that the IRAF RVSAO package used in our analysis adopted $c=299,792.5$~km/s. As pointed out by \citet[][and references therein]{davis19}, further analyses of our data and other redshift surveys should always be done in units of $z$ to avoid various systematic biases. This is especially important for peculiar velocity studies. The median statistical uncertainty of the redshifts was $c\sigma(z)=41$~km/s. We estimated the systematic uncertainty of our measurements by making repeated observations of several bright galaxies that serve as radial velocity standards (N1316, N1365, N1404, N5846, N7552) and comparing our error-weighted mean redshifts with reference values\footnote{See http://www.cfa.harvard.edu/$\sim$dfabricant/huchra/zcat/\\templates.dat}. We found a negligible $c\langle \Delta z\rangle =4\pm14$~km/s, consistent with our typical statistical uncertainty of 23 km/s for these objects.

Table~\ref{tb:cat} presents our results, following the same format and layout as Table 3 of H12. The columns are described in detail in the table footnote and in the machine-readable version available online. There are 1,068 entries (997 new observations, 39 previously-unpublished measurements, 26 reobserved galaxies, 5 measurements from the literature, and 1 object replaced).

The overall 2.4\% incompleteness in redshifts from H12 was far from uniform across the sky. It was heavily biased toward the ZoA, where many prominent nearby large-scale structures are located: the Great Attractor, the Perseus-Pisces and Ophiuchus superclusters, and the Local Void (among others). The incompleteness in H12 increased from 6.3\% at $|b| < 25^\circ$ to 17.8\% at $|b| < 10^\circ$, leaving many of the aforementioned features poorly sampled at low Galactic latitudes. This was due to the inherent difficulties of getting good-quality spectra for galaxies affected by increased dust extinction and higher stellar density and sky brightness levels. Even along the Galactic equator the new redshifts are inhomogeneously distributed. 

This is illustrated in Fig.~\ref{fig:zoa}, with the top panel zoomed into the ZoA and the bottom panel displaying histograms of the new 2MRS observations as a function of Galactic longitude; 90\% (57\%) of these measurements are contained within $|b|=25^\circ\ (10^\circ)$. Many (but not all) of the clumps in new redshifts are associated with known large-scale structure features. This could well be of relevance to the aforementioned studies of the dipole motion and local flow fields, and to studies of group properties, such as those presented in our companion paper \citep{lambert19}.

\begin{figure*}
\begin{center}
  \includegraphics[width=\textheight,angle=90]{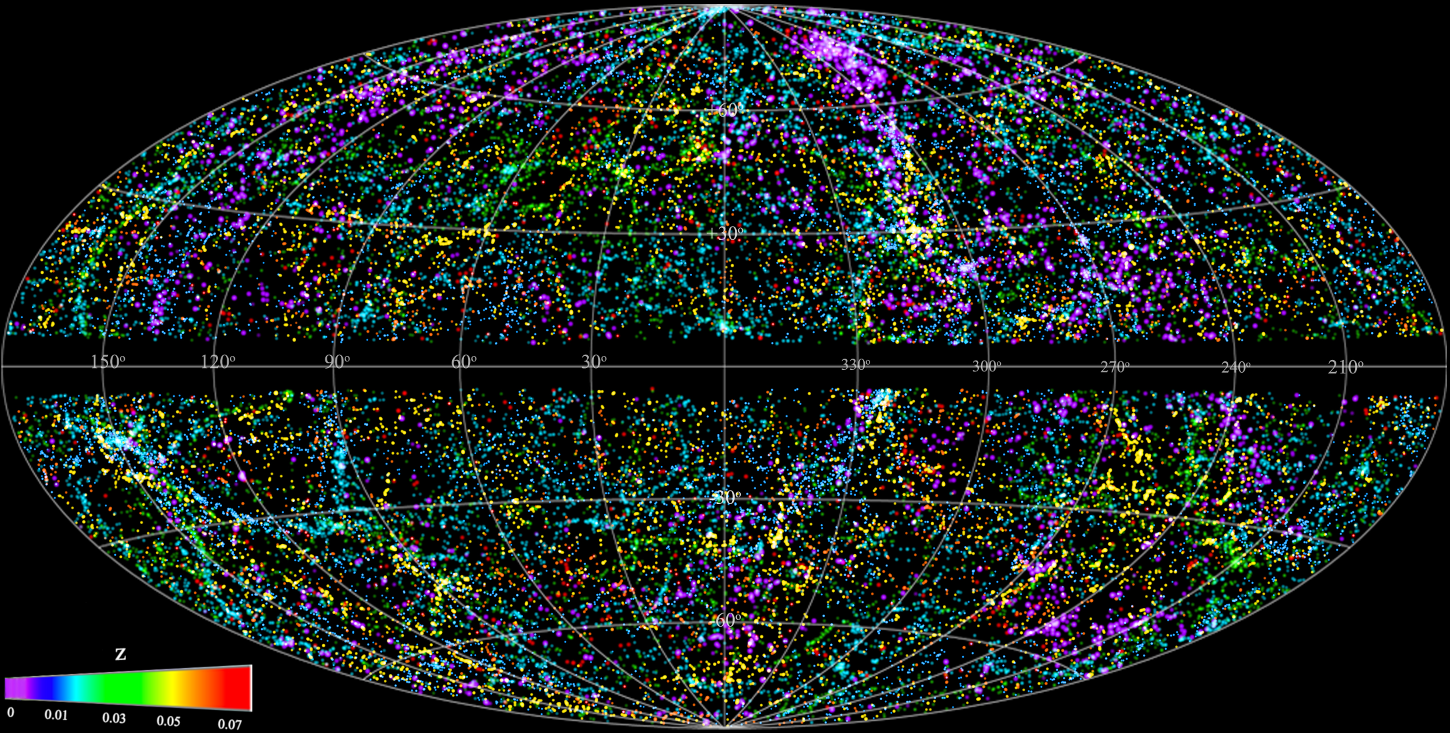}
\end{center}
\caption{All-sky distribution of 2MRS galaxies in Galactic coordinates, using color to convey their redshifts. Completeness in the ZoA (relative to H12) yields a tighter delination of the survey boundaries. \label{fig:col}}
\end{figure*}

\begin{figure*}
 \begin{center}
 \includegraphics[height=0.55\textheight,angle=0]{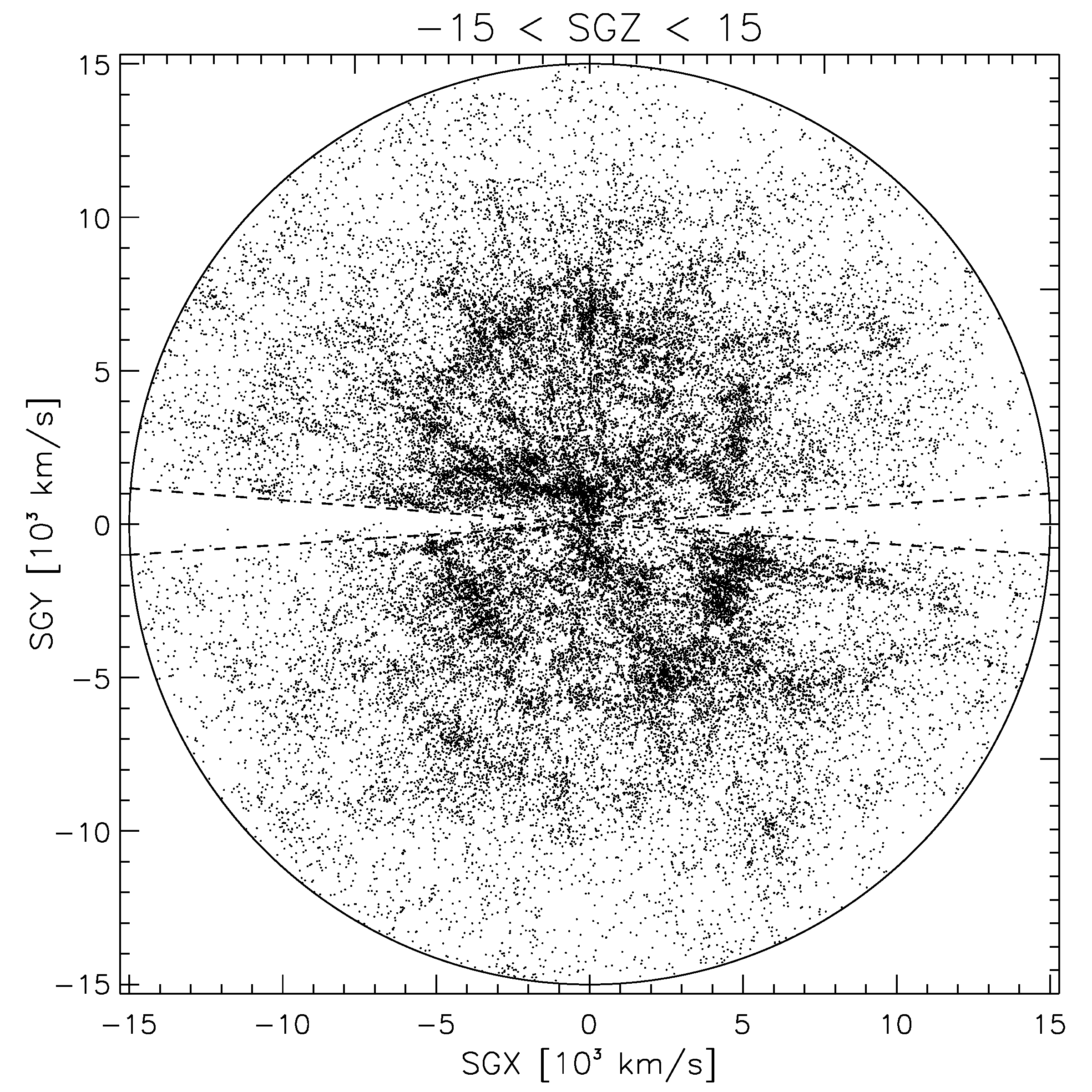}\\
 \includegraphics[height=0.385\textheight,angle=0]{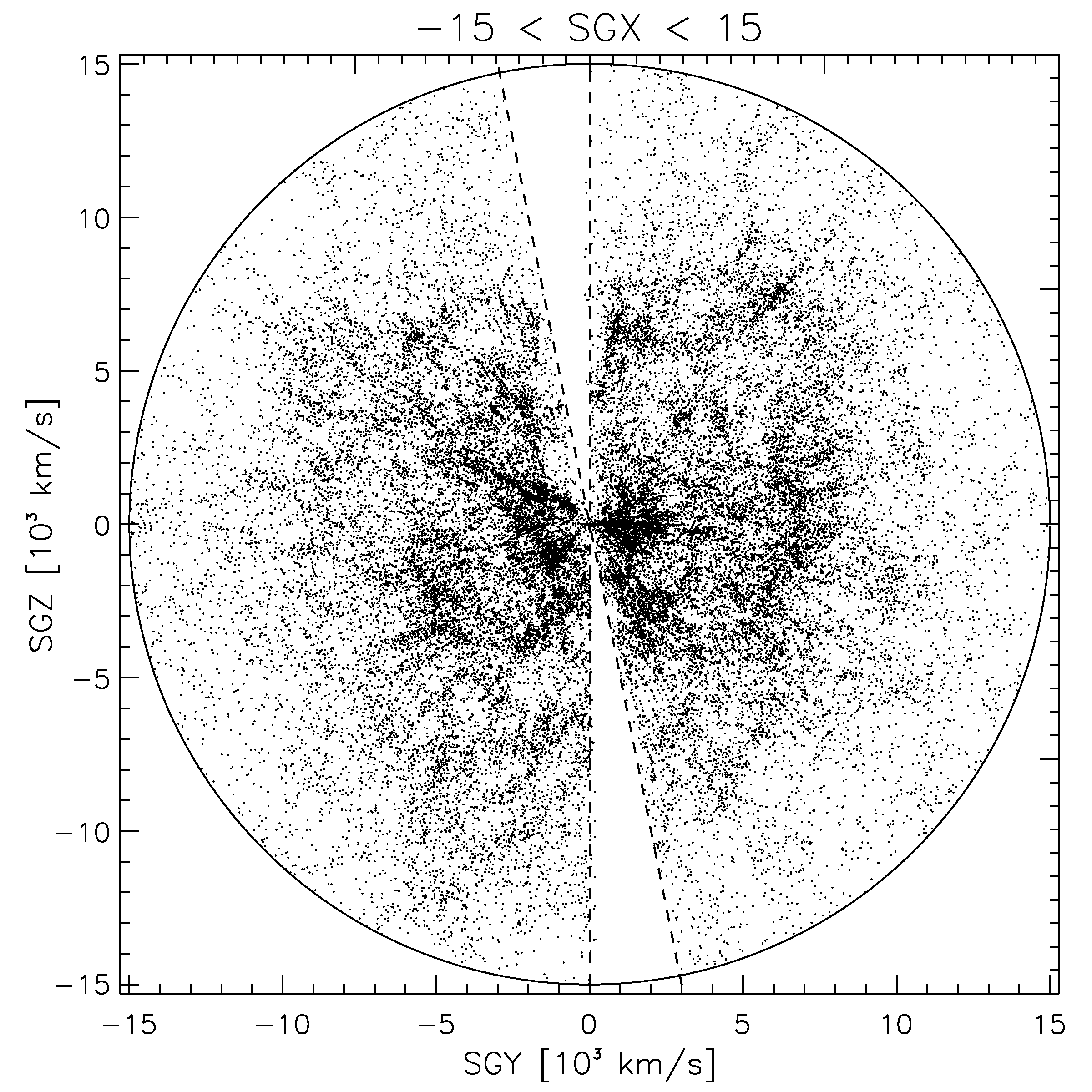}
 \includegraphics[height=0.385\textheight,angle=0]{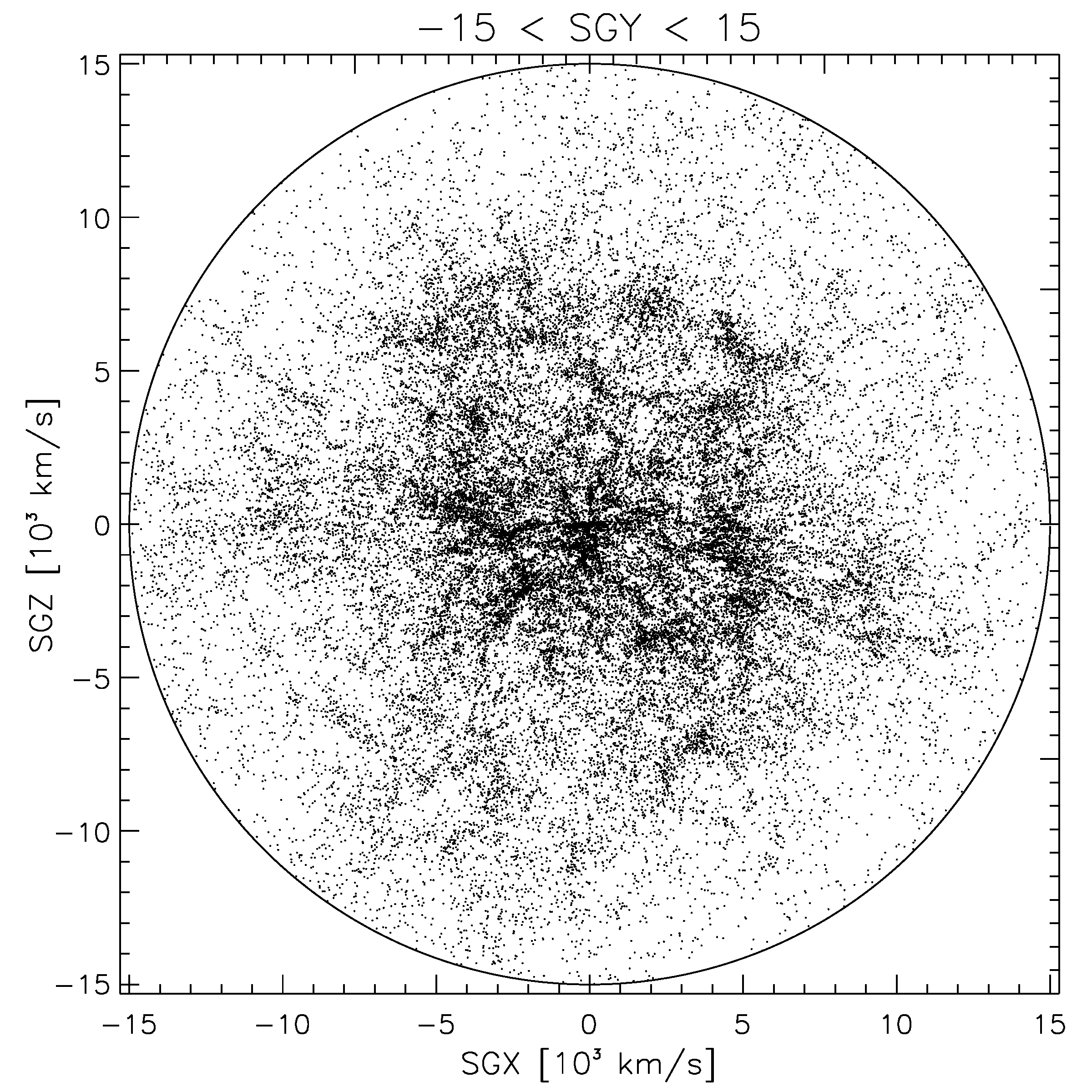}
 \end{center}
\caption{Distribution of 2MRS galaxies in SuperGalactic coordinates, using units of $10^3$~km/s. Top: projection into the (SGX,SGY) plane; bottom left: projection into the (SGY,SGZ) plane; bottom right: projection into the (SGX,SGZ) plane. The approximate boundaries of our survey due to the innermost ZoA are shown using dashed lines.\label{fig:sgl}}
\end{figure*}

In the following, we provide some insight into some of these new features which are also apparent in Fig.~\ref{fig:col} and Fig.~\ref{fig:sgl}. The former shows the all-sky 2MRS galaxy distribution in Galactic coordinates, color-coded for redshift, where the color-coding is closely matches H12. Note how neatly the completed 2MRS lines up with its nominal latitude boundaries compared to the more ragged delineation in H12 around $|b| = 5^\circ$. The latter displays the distribution of 2MRS galaxies in various projections of the SuperGalactic plane.

\begin{deluxetable*}{lrrrrcccccccc}
\tabletypesize{\scriptsize}
\setlength{\tabcolsep}{0.02in}
\tablewidth{0pc} 
\tablecaption{2MRS Catalog (columns 1-13)\label{tb:cat}}
\tablehead{
\colhead{(1)}      & \colhead{(2)}   & \colhead{(3)}  & \colhead{(4)}   & \colhead{(5)}  & \colhead{(6)}   & \colhead{(7)}  & \colhead{(8)}   & \colhead{(9)}    & \colhead{(10)}    & \colhead{(11)}   & \colhead{(12)}          & \colhead{(13)}       \\
\colhead{2MASS ID} & \colhead{R.A.}  &\colhead{Dec.}  & \colhead{$l$}   &\colhead{$b$}   & \colhead{$K^0_s$} &\colhead{$H^0$} & \colhead{$J^0$} &\colhead{$K^0_{s,t}$} & \colhead{$H^0_t$} &\colhead{$J^0_t$} & \colhead{$\sigma(K^0_s)$} &\colhead{$\sigma(H^0)$} \\
\colhead{}         & \multicolumn{2}{c}{(deg)}        & \multicolumn{2}{c}{(deg)}        & \multicolumn{8}{c}{(mag)}}
\startdata
$02485298+5302143$ &  42.22079 &  53.03735 & 140.17172 &  -5.85763 &  9.537 &  9.726 & 10.426 &  9.445 &  9.632 & 10.341 & 0.030 & 0.024 \\
$04270415+2027093$ &  66.76727 &  20.45259 & 176.31357 & -19.38375 &  9.883 & 10.093 & 10.758 &  9.799 &  9.993 & 10.671 & 0.036 & 0.029 \\
$05384231+1544532$ &  84.67630 &  15.74814 & 190.44562 &  -8.25286 &  9.928 & 10.136 & 10.787 &  9.848 & 10.057 & 10.708 & 0.037 & 0.024 \\
$09463018-2134178$ & 146.62576 & -21.57161 & 255.66951 &  23.88194 & 10.101 & 10.358 & 11.044 & 10.000 & 10.251 & 10.931 & 0.048 & 0.035 \\
$19121580-6358361$ & 288.06598 & -63.97664 & 332.13776 & -26.43528 & 10.272 & 10.558 & 11.220 & 10.021 & 10.009 & 10.678 & 0.054 & 0.044 \\
$12525011-1018361$ & 193.20883 & -10.31006 & 303.49725 &  52.56010 & 10.410 & 10.676 & 11.359 & 10.305 & 10.436 & 11.168 & 0.046 & 0.028 \\
$03212772+4048059$ &  50.36548 &  40.80168 & 151.24796 & -13.67875 & 10.462 & 10.723 & 11.401 & 10.267 & 10.485 & 11.152 & 0.045 & 0.036 \\
$04194441+3557293$ &  64.93507 &  35.95812 & 163.36200 & -10.07978 & 10.618 & 10.896 & 11.647 & 10.521 & 10.790 & 11.529 & 0.046 & 0.038 \\
$06074593+3225063$ &  91.94147 &  32.41850 & 179.36438 &   5.86340 & 10.692 & 10.944 & 11.627 & 10.534 & 10.850 & 11.523 & 0.045 & 0.044 \\
$05430236+1620591$ &  85.75992 &  16.34973 & 190.46982 &  -7.05045 & 10.723 & 10.951 & 11.632 & 10.604 & 10.801 & 11.490 & 0.051 & 0.030 \\
$05174145+1936010$ &  79.42268 &  19.60033 & 184.42633 & -10.39923 & 10.726 & 11.122 & 11.994 & 10.629 & 10.989 & 11.887 & 0.050 & 0.036 \\
$06010670+3342353$ &  90.27802 &  33.70980 & 177.55162 &   5.26711 & 10.819 & 11.247 & 12.304 & 10.733 & 11.135 & 12.164 & 0.044 & 0.047 \\
$05323561+1522511$ &  83.14842 &  15.38087 & 189.98283 &  -9.69933 & 10.841 & 11.155 & 11.843 & 10.691 & 11.008 & 11.694 & 0.061 & 0.038 \\
$18570768-7828212$ & 284.28232 & -78.47256 & 315.86765 & -26.82081 & 10.841 & 11.444 & 12.261 & 10.825 & 11.237 & 11.943 & 0.050 & 0.052 \\
$04301670+2326448$ &  67.56960 &  23.44573 & 174.42603 & -16.87470 & 10.857 & 11.029 & 11.807 & 10.698 & 10.819 & 11.617 & 0.063 & 0.039 \\
$05242105+1422371$ &  81.08781 &  14.37709 & 189.77205 & -11.91985 & 10.905 & 11.138 & 11.841 & 10.816 & 11.069 & 11.796 & 0.056 & 0.033 \\
$05232510+1633198$ &  80.85459 &  16.55557 & 187.77589 & -10.93586 & 10.916 & 11.113 & 11.789 & 10.823 & 11.010 & 11.705 & 0.062 & 0.035 \\
$17413903-0437191$ & 265.41263 &  -4.62198 &  20.59406 &  13.25158 & 10.970 & 11.223 & 11.972 & 10.714 & 10.943 & 11.717 & 0.052 & 0.045 \\
$08324123-5113345$ & 128.17188 & -51.22622 & 268.17825 &  -6.73042 & 10.995 & 11.237 & 12.090 & 10.776 & 11.035 & 11.920 & 0.069 & 0.051 \\
$07365795-4047484$ & 114.24139 & -40.79684 & 254.12871 &  -9.49531 & 11.005 & 11.153 & 11.870 & 10.886 & 10.960 & 11.604 & 0.070 & 0.041 \\
$07244714+1405193$ & 111.19660 &  14.08878 & 203.90164 &  13.66018 & 11.031 & 11.368 & 12.270 & 10.968 & 11.152 & 12.082 & 0.039 & 0.030 \\
$18044513+1731392$ & 271.18802 &  17.52752 &  43.84589 &  17.99977 & 11.102 & 11.377 & 12.122 & 10.922 & 11.194 & 11.956 & 0.057 & 0.042 \\
$05272524+1612051$ &  81.85517 &  16.20144 & 188.60564 & -10.31789 & 11.107 & 11.403 & 12.143 & 11.019 & 11.311 & 12.080 & 0.059 & 0.036 \\
$20595692-5533431$ & 314.98718 & -55.56196 & 341.67456 & -40.07023 & 11.121 & 11.324 & 11.908 & 10.919 & 11.111 & 11.734 & 0.083 & 0.055 \\
$14593308-5154213$ & 224.88797 & -51.90596 & 322.14438 &   6.12193 & 11.123 & 11.367 & 12.108 & 10.687 & 11.359 & 11.769 & 0.090 & 0.077 \\
$23122811-6131165$ & 348.11710 & -61.52133 & 321.77420 & -51.83395 & 11.123 & 11.450 & 12.069 & 11.038 & 11.348 & 11.944 & 0.058 & 0.044 
\enddata
\tablecomments{This table is presented in its entirety in the online version of the paper.}
\end{deluxetable*}

\addtocounter{table}{-1}
\begin{deluxetable*}{lcccclllllrrrr}
\tabletypesize{\scriptsize}
\tablewidth{0pc} 
\tablecaption{2MRS Catalog (columns 14-26)}
\tablehead{
\colhead{(1)}      & \colhead{(14)}          & \colhead{(15)}            & \colhead{(16)}            & \colhead{(17)}           & \colhead{(18)}    & \colhead{(19)}        & \colhead{(20)}         & \colhead{(21)} & \colhead{(22)}  & \colhead{(23)} & \colhead{(24)}   & \colhead{(25)}   & \colhead{(26)}        \\
\colhead{2MASS ID} & \colhead{$\sigma(J^0)$} & \colhead{$\sigma(K^0_t)$} & \colhead{$\sigma(H^0_t)$} & \colhead{$\sigma(J^0_t)$} & \colhead{$E_{BV}$} &\colhead{$r_{\rm{iso}}$} & \colhead{$r_{\rm{ext}}$} &\colhead{$b/a$} & \colhead{flags} & \colhead{Type} & \colhead{t\_src} & \colhead{$cz$}    & \colhead{$c\sigma(z)$} \\
\colhead{}         & \multicolumn{4}{c}{(mag)}                                                                                  & \colhead{(mag)}   &\multicolumn{2}{c}{($\log_{10} \arcsec$)}        &\colhead{}      & \colhead{}      &\colhead{}      & \colhead{}       & \multicolumn{2}{c}{(km/s)}}      
\startdata
$02485298+5302143$ & 0.020 & 0.029 & 0.024 & 0.022 & 0.493 & 1.615 & 1.776 & 0.420 & 0111 &  3    & ZC &  4735 &  25 \\
$04270415+2027093$ & 0.023 & 0.033 & 0.026 & 0.026 & 0.576 & 1.529 & 1.702 & 0.460 & 0000 &  0X   & ZC &  3856 &  21 \\
$05384231+1544532$ & 0.021 & 0.034 & 0.022 & 0.022 & 0.518 & 1.504 & 1.680 & 0.560 & 0000 & 98    & NN &  5234 &  26 \\
$09463018-2134178$ & 0.024 & 0.046 & 0.032 & 0.027 & 0.047 & 1.444 & 1.634 & 0.780 & 0666 &  0    & JH &  8794 &  31 \\
$19121580-6358361$ & 0.030 & 0.047 & 0.031 & 0.023 & 0.041 & 1.455 & 1.664 & 0.500 & 0000 &  0    & JH & 11045 &  44 \\
$12525011-1018361$ & 0.021 & 0.046 & 0.027 & 0.028 & 0.041 & 1.449 & 1.692 & 0.300 & 0333 &  4    & JH &  4096 &  18 \\
$03212772+4048059$ & 0.026 & 0.043 & 0.035 & 0.033 & 0.149 & 1.330 & 1.604 & 0.720 & 0331 & -2    & ZC &  6088 &  21 \\
$04194441+3557293$ & 0.030 & 0.042 & 0.038 & 0.034 & 0.588 & 1.356 & 1.556 & 0.420 & 0333 &  3    & ZC &  5248 &  22 \\
$06074593+3225063$ & 0.033 & 0.040 & 0.045 & 0.039 & 0.715 & 1.398 & 1.617 & 0.500 & 0000 & 98    & NN &  8608 &  42 \\
$05430236+1620591$ & 0.024 & 0.044 & 0.029 & 0.026 & 0.485 & 1.303 & 1.512 & 0.620 & 0000 & 98    & NN &  5737 &  27 \\
$05174145+1936010$ & 0.035 & 0.046 & 0.039 & 0.040 & 0.580 & 1.520 & 1.695 & 0.180 & 0133 &  2    & ZC &  5272 &  40 \\
$06010670+3342353$ & 0.053 & 0.046 & 0.049 & 0.062 & 0.799 & 1.394 & 1.596 & 0.220 & 0000 & 98    & NN &  7698 &  28 \\
$05323561+1522511$ & 0.036 & 0.052 & 0.038 & 0.039 & 0.641 & 1.338 & 1.543 & 0.640 & 0000 & 98    & NN &  6402 &  11 \\
$18570768-7828212$ & 0.038 & 0.058 & 0.059 & 0.044 & 0.156 & 1.161 & 1.502 & 0.560 & 0000 &  0B   & JH & 12977 &  46 \\
$04301670+2326448$ & 0.036 & 0.056 & 0.038 & 0.040 & 0.900 & 1.297 & 1.526 & 0.600 & 0333 &  5    & JH &  5424 &  44 \\
$05242105+1422371$ & 0.029 & 0.049 & 0.034 & 0.033 & 0.419 & 1.356 & 1.557 & 0.360 & 0444 &  2    & ZC &  5882 &  23 \\
$05232510+1633198$ & 0.030 & 0.054 & 0.035 & 0.040 & 0.403 & 1.332 & 1.537 & 0.520 & 0000 &  0X   & ZC &  5710 &  36 \\
$17413903-0437191$ & 0.044 & 0.080 & 0.066 & 0.066 & 0.717 & 1.422 & 1.707 & 0.500 & 0111 &  7    & ZC &  7962 &  14 \\
$08324123-5113345$ & 0.056 & 0.064 & 0.056 & 0.072 & 0.960 & 1.243 & 1.526 & 0.720 & 0000 & 98    & NN & 11521 &  84 \\
$07365795-4047484$ & 0.037 & 0.064 & 0.039 & 0.037 & 0.672 & 1.179 & 1.419 & 0.720 & 0333 & 98    & NN &  8725 &  57 \\
$07244714+1405193$ & 0.027 & 0.036 & 0.028 & 0.030 & 0.126 & 1.017 & 1.301 & 0.960 & 0333 &  4    & JH &  5405 &  29 \\
$18044513+1731392$ & 0.031 & 0.051 & 0.045 & 0.037 & 0.120 & 1.281 & 1.542 & 0.540 & 0333 & -1    & JH &  5544 &  16 \\
$05272524+1612051$ & 0.036 & 0.055 & 0.038 & 0.043 & 0.615 & 1.318 & 1.525 & 0.420 & 0111 &  2    & ZC &  5721 &  33 \\
$20595692-5533431$ & 0.036 & 0.070 & 0.054 & 0.038 & 0.049 & 1.428 & 1.638 & 0.440 & 0000 &  0    & JH &  2003 &  23 \\
$14593308-5154213$ & 0.074 & 0.114 & 0.166 & 0.130 & 0.584 & 1.270 & 1.708 & 0.780 & 0222 & 98    & NN & 11102 &  32 \\
$23122811-6131165$ & 0.029 & 0.053 & 0.045 & 0.032 & 0.018 & 1.204 & 1.439 & 0.880 & 0000 & -2    & JH &  7884 &  29 
\enddata
\tablecomments{This table is presented in its entirety in the online version of the paper.}
\end{deluxetable*}

\addtocounter{table}{-1}
\begin{deluxetable*}{lcll}
\tabletypesize{\scriptsize}
\tablewidth{0pc} 
\tablecaption{2MRS Catalog (columns 27-29)}
\tablehead{
\colhead{(1)}      & \colhead{(27)} & \colhead{(28)}   & \colhead{(29)}      \\
\colhead{2MASS ID} & \colhead{cat}  & \colhead{v\_src} & \colhead{Catalog ID}}
\startdata
$02485298+5302143$ & F & 20192MRS.FLWO.0000M & 02485298+5302143\\
$04270415+2027093$ & F & 20192MRS.FLWO.0000M & 04270415+2027093\\
$05384231+1544532$ & F & 20192MRS.FLWO.0000M & 05384231+1544532\\
$09463018-2134178$ & Z & 20192MRS.SAAO.0000M & 09463018-2134178\\
$19121580-6358361$ & Z & 20192MRS.SAAO.0000M & 19121580-6358361\\
$12525011-1018361$ & L & 20192MRS.CSLO.0000M & 12525011-1018361\\
$03212772+4048059$ & F & 20192MRS.FLWO.0000M & 03212772+4048059\\
$04194441+3557293$ & F & 20192MRS.FLWO.0000M & 04194441+3557293\\
$06074593+3225063$ & F & 20192MRS.FLWO.0000M & 06074593+3225063\\
$05430236+1620591$ & F & 20192MRS.FLWO.0000M & 05430236+1620591\\
$05174145+1936010$ & F & 20192MRS.FLWO.0000M & 05174145+1936010\\
$06010670+3342353$ & F & 20192MRS.FLWO.0000M & 06010670+3342353\\
$05323561+1522511$ & F & 20192MRS.FLWO.0000M & 05323561+1522511\\
$18570768-7828212$ & Z & 20192MRS.SAAO.0000M & 18570768-7828212\\
$04301670+2326448$ & F & 20192MRS.FLWO.0000M & 04301670+2326448\\
$05242105+1422371$ & F & 20192MRS.FLWO.0000M & 05242105+1422371\\
$05232510+1633198$ & F & 20192MRS.FLWO.0000M & 05232510+1633198\\
$17413903-0437191$ & F & 20192MRS.FLWO.0000M & 17413903-0437191\\
$08324123-5113345$ & L & 20192MRS.CSLO.0000M & 08324123-5113345\\
$07365795-4047484$ & L & 20192MRS.CSLO.0000M & 07365795-4047484\\
$07244714+1405193$ & L & 20192MRS.CSLO.0000M & 07244714+1405193\\
$18044513+1731392$ & F & 20192MRS.FLWO.0000M & 18044513+1731392\\
$05272524+1612051$ & L & 20192MRS.CSLO.0000M & 05272524+1612051\\
$20595692-5533431$ & Z & 20192MRS.SAAO.0000M & 20595692-5533431\\
$14593308-5154213$ & Z & 20192MRS.SAAO.0000M & 14593308-5154213\\
$23122811-6131165$ & Z & 20192MRS.SAAO.0000M & 23122811-6131165
\enddata
\tablecomments{This table is presented in its entirety in the online version of the paper. Codes for column 27 that identify our new observations are: [F], FLWO/FAST; [G], SOAR/Goodman; [L], CASLEO/REOSC; [M], MDM/OSMOS; [Z], SAAO/CassSpec or SpUpNIC. [N] denotes redshifts from NED (NASA/ADS bibliographic code given in column 28). [2], [6], [K], [O], [P], [S] refer to previously unpublished observations by RKK (see online table for details).}
\end{deluxetable*}

\clearpage

In the Perseus-Pisces complex (PP; $150^\circ < l < 200^\circ$) a filamentary extension from Perseus ($l \sim 150^\circ$) toward the Galactic Plane around $165^\circ$ shows up quite prominently for the first time. The feature connects to a ridge in the ZoA, which encompasses the 3C 129 cluster that links PP to Lynx \citep[see][]{ramatsoku16,kraan18}. It also traces the continuation of the second PP arm ($l \sim 90^\circ$) closer to the ZoA, connecting to a filamentary extension across the ZoA found with \ion{H}{1} observations \citep{kraan18}. The concentration around $l\sim~100^\circ, b\sim -5^\circ$ (seen in green in Fig.~\ref{fig:col}) is a complete surprise, seemingly forming an isolated group or small cluster around 12,000~km/s. The clump of redshifts at $l\sim 0^\circ, b\sim +5^\circ$ (seen in cyan in Fig.~\ref{fig:col}) is centered around 10,000~km/s and hence clearly part of the Ophiuchus supercluster \citep{wakamatsu05}.

The number of new redshifts in the general direction of the Great Attractor (GA; $l=320^\circ - 330^\circ$) is also quite high. While quite a few of the new redshifts are at the GA distance and are partly linked to the Norma cluster \citep{woudt08}, the majority actually peak around 12-16,000~km/s and are, therefore, more likely to be associated with the overdensity encompassing the Ara and TriAus clusters in that region. Two further clumps around $l=250^\circ$ and $270^\circ$ also rise up in the histogram of Fig~\ref{fig:zoa}, both with a high concentration of galaxies around 12-13,000~km/s. The latter also has a significant number of galaxies that lie around 18-20,000~km/s and thus seems to form part of the Vela supercluster \citep{kraan17}.

\acknowledgments

We thank the dedicated staff of FLWO, SAAO, CASLEO, SOAR and MDM which made these observations possible, and to the respective telescope allocation committees for their support over the past two decades. We dedicate this paper to the memory of John Huchra -- a friend, colleague, and mentor who left us too soon. Typing services provided by Fang, Inc.

\facilities{FLWO:1.5m (FAST), Radcliffe (SpUpNIC), CASLEO:JST (REOSC), SOAR (Goodman), Hiltner (OSMOS)}

\bibliographystyle{aasjournal}
\bibliography{macri}

\appendix
\setcounter{table}{0}
\renewcommand{\thetable}{A\arabic{table}}
\section{Changes to the 2MRS Catalog}
  
In the table below we list the deletions of entries in the 2MRS catalog, relative to H12. We provide redshifts when available (in some cases, the object was rejected by visual inspection before obtaining a spectrum).

\ \par

\begin{deluxetable*}{ll}[b]
\tablewidth{0pc} 
\tablecaption{Deletions from the 2MRS Catalog\label{tab:chg}}
\tablehead{
\colhead{2MASS ID} & \colhead{Reason for deletion}}
\startdata
$01515799+7126330$ & star, $v=-59\pm46$ \\
$04253337+4201067$ & star, $v=-66\pm44$ \\
$04304237+6347222$ & star \\
$05504348+1027466$ & star+galaxy, $v=-90\pm72$ and $cz=12232\pm78$ \\
$06073397-0541583$ & star  \\
$06081453-0934490$ & stsar, $v=51\pm12$ \\
$06091575-1011338$ & star \\
$06341666-0915552$ & star, $v=126\pm12$ (H$\alpha$) \\
$06355814+2240350$ & duplicate of $06355890+2240352$, redshift already in H12 \\
$08290119-5121097$ & star+galaxy \\
$09002997-2224422$ & star+galaxy \\
$09295260-6210339$ & duplicate of $09295846-6210599$ \\
$13224883-5629132$ & star+galaxy, $v=-37\pm54$\\
$14201250-6821596$ & stars \\
$14302060-5355016$ & galactic source? WISE W4=2.3 mag\\
$14522589-6711268$ & star+galaxy, $v=371\pm43$ \\
$15341907-4534416$ & star+galaxy, $v=97\pm51$\\
$15524980-4032516$ & star+galaxy, $v=187\pm39$\\
$16123697-3629426$ & star+galaxy, $v=36\pm47$\\
$16131380-6234545$ & star+galaxy, $v=147\pm68$\\
$16292221-2652276$ & star\\
$16523424-1959028$ & star+galaxy, 6dF $v=-56\pm45$ \\
$16584354-6629079$ & star \\
$19190956-0258564$ & star+galaxy, $v=487\pm141$\\
$20015711-0649486$ & star+galaxy, $v=-108\pm26$ and $cz=1469\pm91$ \\
$20451931-3514274$ & duplicate of $20451841-3514264$ \\
$21504406+4159299$ & star+galaxy, $v=-183\pm55$ and $cz=5572\pm54$ \\
\enddata
\end{deluxetable*}
\ \par
\clearpage
\end{document}